# Enable an Open Software Defined Mobility Ecosystem through VEC-OF

Sanchu Han
Standard, Industry Development
Futurewei Technologies Inc.
Seattle, USA
sanchu.han@futurewei.com

Yong He
Software R&D Lab
Futurewei Technologies Inc.
Dallas, USA
yong.he@futurewei.com

Yin Ding
Cloud R&D Lab
Futurewei Technologies Inc.
Seattle, USA
yin.ding@futurewei.com

*Abstract*—OEMs and new entrants can take the Mobility as a Service market (MaaS) as the entry point, upgrade its E/E (Electric and Electronic) architecture to be C/C (Computing and Communication) architecture, build one open software defined and data driven software platform for its production and service model, use efficient and collaborative ways of vehicles, roads, cloud and network to continuously improve core technologies such as autonomous driving, provide MaaS operators with an affordable and agile platform. In this paper we present one new framework, VEC-OF (Vehicle-Edge-Cloud Open Framework), which is a new data and AI centric vehicle software framework enabling a much safer, more efficient, connected and trusted MaaS through cooperative vehicle, infrastructure and cloud capabilities and intelligences.

*Keywords—MaaS, CC, SDM, AUTOSAR, AD, ECU, VEC-OF*

## I. INTRODUCTION

With new business demands like MaaS and applications such as AD (Autonomous driving) and intelligent cockpit etc., the vehicle industry is facing a rapid transformation to be digitalized and software defined.

In order to satisfy those new requirements there are two critical tasks. First, the E/E architecture needs to be upgraded to a C/C architecture in Figure 1. Under this new architecture E/E evolves to be a vehicle or zone domain centralized, uses the SOA (Service Oriented or micro-service Architecture) concept to supply the upper layer with secure "control" APIs. Then, microcontroller-based embedded systems for ECUs are designed as a C/C centralized hardware platform.

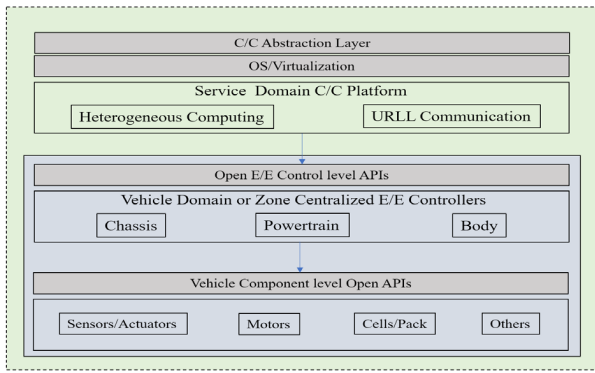

Figure 1- From E/E Architecture to C/C Architecture

Second, on top of the C/C architecture there should be one open data-centric software infrastructure layer for different domains' applications and ecosystems across vehicle, road and cloud, this is shown in Figure 2.

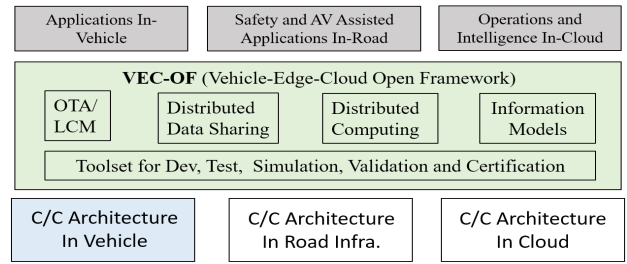

Figure 2- Open Software Framework Function Diagram

While most OEMs' E/E architectures are still in "distributed" fashion, a couple of leading OEMs may enter the initial phase of "domain centralized". For the software industry, the widely accepted architecture is AUTOSAR (Automotive Open System Architecture). However, both CP (Classical platform) and AP (Adaptive platform) of AUTOSAR are still "function cluster" oriented. Some critical capabilities are not clearly addressed in the architecture.

In this paper we will mainly focus on the software infrastructure working with C/C architecture, VEC-OF, one open software framework handling the software management, critical computing task allocation, data flow handling in the vehicle and information sharing across the vehicle, road and cloud. For example:

1. In vehicle: one abstract software layer in C/C to simplify multi-vendors' integration for vehicle controls including dynamic sensors and actuators, AD/ADAS sensors etc.

2. Across vehicle: one common software framework, which is independent from its underlying communication protocols and upper level applications, to facilitate the multi-industry collaborations cross automotive, ITS (Intelligent transportation system) and cloud systems.

## II. NEW BUSINESS REQUIREMENTS AND TECHNOLOGY TRENDS

### A. User experience and new value creation drive SDV (Software Defined Vehicle) through connected capabilities and cooperative intelligence

Traditionally after a vehicle is delivered to the customer the life cycle of the vehicle is basically transferred to the car dealership for the vehicle's aftermarket maintenance. In case OEMs need to recall the vehicles due to some issue, the vehicles have to be returned back to the dealership and OEMs cover all the costs. The Human Machine Interface of the vehicle is not experience-centric designed, most of the non-safety related features are hard to use, they are basically "dead" and have never been willingly updated by customers. While some industry innovators remotely update their application software and even some performance related software through OTA quite frequently as smart phone vendors do. They are also taking this capability to enable new business potentials with the customers through the life cycle of the vehicle.

With 5G, the global communication standard and vertical industry service enabler, the standalone vehicle intelligence can be greatly extended through connected intelligence from road and cloud for content services, cooperative autonomous driving and efficient transportation in a much cost effective way.

### B. The data is the new "fuel" of the vehicle

Some innovations are data centric SDV beyond the baseline of EV (Electric Vehicle). For example, in CVPR 2020 workshop on "Scalability in Autonomous Driving" Andrej Karpathy from Tesla shows how Tesla leverages its fleet data for "Stop sign" detection in different scenes. MaaS operators can leverage data for customer experience, operations, regulations, and even research and development for AI algorithms. Billions of miles of traffic environment and behavioral data can be used for its autonomous driving testing and on-line learning. OEMs without a data strategy in their product and operations will be outdated in the near future just like feature phone vendors were outdated by smartphones.

Structured and formatted data matters for the value creation. In a vehicle the data from different sensors need to follow specified format for the downstream application processing. Additionally data across different systems including ITS and cloud needs to reach the consensus on the exchange formats and underlying communication standards as well.

### C. AI centric applications are driving the in-vehicle heterogeneous computing requirements

More and more applications in ADAS (Advanced driver assist system)/ADS (Autonomous driving system), intelligent cockpit and HMI systems etc. are applying ML and DL algorithms. Those data intensive driven scenarios are dramatically reshaping the computing architecture of MCU/ECU to be computer based heterogeneous computing architecture, where for both training and inference cases the computing tasks will be potentially orchestrated among CPU, GPU and NPU etc. The computing framework of the vehicle needs to take this into consideration when the vehicle is evolving more intelligently down the road.

### D. Mobility service has the strongest business needs to drive more modular, reusable, replaceable hardware and upgradable software

We can see a clear business desire for a software defined mobility from MaaS operators. The TCO (Total Cost Ownership) is the number one factor for them and the main components are drivers and vehicles.

- AD is projected to remove the driver cost but it will take a long time to achieve that goal. AD in low-to-mid speed for level 3 and beyond are commercially achievable through closed, fixed or more controllable operational design domains such as city bus, sea ports and airport luggage shipping etc. They are unlikely to be achieved for ownership cars due to high cost equipped sensors of current technologies. The AV capability can be improved through the mobility service using incremental software updates. In the short to medium term, AD for MaaS is not going to completely replace drivers, instead it will aim at greatly increasing traffic safety and efficiency.

- The cost reduction of the fleet plus maintenance can be achieved through i) open modular, rechargeable and replaceable parts like motors and battery which can dramatically increase the operating life time and utilization for the vehicle. Furthermore, it will drive the need of standardization for the vehicle parts. ii) centralized C/C architecture through consolidated E/E by reducing the numbers of ECUs, therefore also simplifying the overall architecture complexities.

### E. Open vehicle Platform enables the ecosystem of software defined mobility

The "function cluster" mode has been defined by the current auto supply chains, it is slow changing as it is adapted to the 4-6 years' vehicle production cycle used by most traditional OEMs. In this mode the function oriented features are being vertically developed and integrated with some amount of effort for inter-ECU communication and integration. There is no platform concept like public cloud and OTT business areas, where one common platform plays the role of the application ecosystem hub for both development and marketplace service.

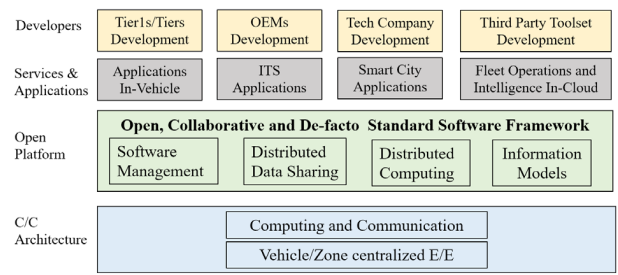

Figure 3- Open platform for new ecosystem

In figure 3 we envision one open platform for the connected vehicles to establish one new ecosystem, which can greatly increase the collaboration efficiency and simplify the integration efforts among multiple industries, while enabling a more rich development ecosystem.

## III. SOFTWARE DEFINED MOBILITY DRIVES FOR A NEW VEC-OF ARCHITECTURE

The automotive industry is undergoing a digital transformation driven by a number of new trends, including connect cars, autonomous vehicles, share-riding and electrification. A new C/C architecture that includes both hardware and software is expected to change very rapidly in the future to keep up with these fast changing industry.

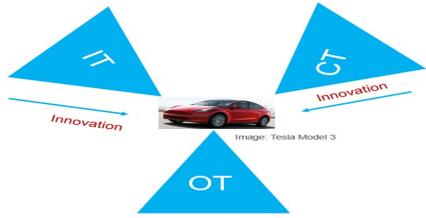

Figure 4 OT-IT-CT Convergence

Ttraditionally the vehicle electronics are OT (Operation Technology) based, through deeply embedded software to control sensors / actuators (figure 4). This OT based approach results in legacy vehicle's mechanical and functional E/E architecture, which is highly decentralized with an average of 50 to 70 ECUs (# of ECUs depends on functional features provided by the Vehicle). The cost control within the existing supply chain is very tight and barely leaves budget for new features during the vehicle's lifespan.

AV has been a hot topic for years. There are a few open source projects, like AUTOWARE [4] and APOLLO [5]. The basic architectures for these projects follow the one shown in Figure 5. They are having some momentum in AV field, however, still have some critical challenges to overcome:

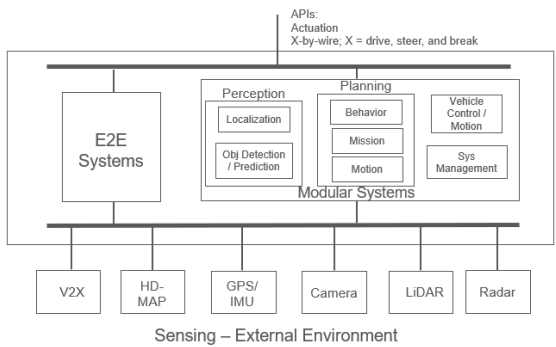

Figure 5 Basic Autonomous Driving Functional Architecture

### A. High cost sensing and computing hardware

It costs thousands of dollar in sensing and computing to achieve L3-L4 AV (not to mention of the immaturity of current AI algorithms). From a financial stand point this is cost prohibiting. However, in case of mobility service for goods and people the new market is much less sensitive to the cost as long as we can leverage much simpler C/C architecture through EV to achieve optimal TCO for the operators. For the MaaS the customers care more about safety, efficiency, and experience instead of branding and social status so that the EV can be configured through a much simpler setting with less ECUs. It is going to drive the MaaS operators a stronger desire to make the vehicle E/E architecture centralized to be software defined. Not only will this further reduce the cost of the vehicle manufacturing, but also dramatically increase the service agility.

### B. Vehicle intelligence without edge and cloud intelligence being supported through the architecture

- Too many corner cases need to be handled by perception algorithms, and furthermore the path planning module it's still an open research topic for AI algorithms in order to make the right decisions on the behavior prediction and actions of complex driving scenarios.

- MaaS for people and goods can be incrementally deployed in more controlled, fixed routes and predicable scenarios. The Vehicle intelligence limitation due to the occlusion and environment can be resolved more cost effectively by a cooperative intelligence from the vehicle transportation infrastructure and cloud.

In short there is no existing framework to take the mobility native requirements into the architecture with a cooperative intelligence and data sharing across vehicle, edge infrastructure and cloud which we would argue will result in a much safer, affordable, feasible and efficient mobility service.

## IV. VEC-OF

In this section we present the VEC-OF, an infrastructure software framework aimed at enabling a rich open ecosystem for SDM. In Figure 6 shows a design time and a runtime architecture. The design time consists of four modules: onboarding of algorithms and applications, service design and deployment for the fleet, traffic generation and testing, scene generation and validation. The runtime can be used in both production and testing environments, and the major components include a service orchestrator to handle the runtime request, a LCM module to manage the fleet through OTA, one distributed heterogeneous computing framework to allocate the computing tasks in the vehicle and the cloud, a geographic distributed data sharing module to ship the data across vehicle, edge and cloud for operations, cooperative autonomous driving and regulations.

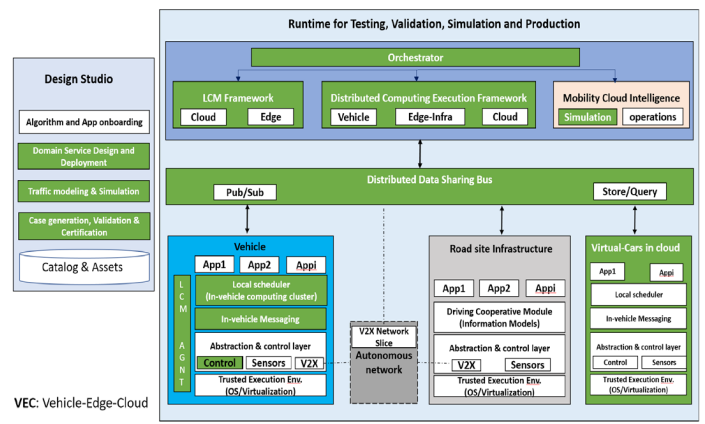

Figure 6- VEC-OF Overall High Level Architecture

### A. OTA-LCM

Today, in the data center cloud infrastructure, KUBERNETES has become the de-facto standard of enterprises for the orchestration. To extend native containerized application orchestration from center data center or public cloud to edges,

KubeEdge [6], another open source CNCF sandbox project, builds upon KUBERNETES and acts an orchestrator in edge cloud infrastructure.

VEC-OF builds upon KubeEdge and KUBERNETES open source projects, provides core infrastructure including a service application orchestrator, Life Cycle Management (LCM) framework, and LCM agents supporting resource management, device and ECUs management in a vehicle resource constrained environment.

- Orchestrator

As the name infers, the orchestrator orchestrates workloads for maximum availability across the whole system. Orchestration means to lay out the patterns of the applications in order for them to work together with other applications. The orchestrator's job is to maintain the application's health, manage resources (computing, networking and storage), and ensure the system resources are in the desired state with efficient utilization.

The orchestrator provides:

- Service discovery: VEC is a sophisticated system, lots of services are registered in the system. The orchestrator provides a server-side service discovery.
- Resilience: maintains desired state throughout the lifespan of the application, restarts / replaces / reschedules individual workload instead of replicating the entire application.
- Continuity: integrates with continuous integration and continuous delivery (CI/CD) to break easily automated comprehend deployment into smaller, manageable steps, and provide automated rollouts and rollbacks.
- Scalability: orchestrates workloads in highly distributed and heterogeneous system

- LCM

The Life Cycle Management framework include resource management in cloud, edge and vehicle, and OTA deployment and update.

The orchestrator behind the LCM provides required services and inventories storage and catalog. It manages services and applications running in the cloud and the edge (a vehicle can be treated as one edge node), provides services discovery, continuity, resiliency and salability.

The LCM agent in vehicle works with the local scheduler managing the local resources for the vehicle constrained resource, and schedules workloads within a vehicle. And through a device twin, the LCM agent builds a virtual mapping of IoT device metadata on the application platform allowing communication between devices, sensors and applications. The LCM agent also provides a communication's channel to the edge and cloud, allowing it to manage devices in the cloud and synchronize device states between vehicle, edge and cloud.

LCM agent runs autonomously during disconnection from edge / cloud. Once reconnected with cloud, it synchronizes the vehicle and device states.

- OTA

The LCM provides OTA updates, simplifies the way of deploying and upgrading firmware, applications and infrastructure.

- Application: the application deployment and upgrade to the vehicle follows the same pattern as KUBERNETES does in the cloud. The orchestrator issues the command, the LCM agent pulls the desired images to the vehicle and deploy them. If anything fails or does not act in the desired manner, it automatically rolls back to the previous version.
- Firmware: The LCM agent evaluates the device to ensure that it's in a healthy state before it allows an upgrade to begin. Then the agent pulls images from the cloud, and update device firmware. Similar to an application upgrade, the LCM agent monitors the health of the device and rolls back to a backup firmware version if there is a failure.

### B. DDSF- Distributed Data Sharing Framework

Intelligence comes from data. Recent developments in data-driven ML/DL algorithms have enabled many use cases and vehicle applications in ADAS/ADS, including remote vehicle telemetry diagnostics etc. However, there are still many unanswered questions, most significantly around data exchanges, data sharing and data queries across vehicle, edge and cloud.

To solve this problem, we propose the DDSF as one fundamental module for VEC-OF: a logically centralized/unified, physically distributed data platform that allows the vehicle, edge and cloud to share data across boundaries while protecting privacy and security.

The design of DDSF is based on one open source project called *zenoh* [7]. It is an Eclipse Edge Native project, which unifies data in motion, data in-use, data at rest and computations. It meets the requirements of traditional pub/sub with geo-distributed storages, queries and computations, while retaining a level of time and space efficiency.

*1) Conceptual Model and APIs*

A data-centric abstraction is provided for applications to read, write, store, query and compute data autonomously and asynchronously. The data read, written, stored, queried and computed by the applications is associated with one or more resources identified by a URI. The examples of URIs:

```
/location/city/road/east-west/traffic_light
/operator-1/fleet/city/**/engine/status
/operator-1/fleet/city/vehicle/hd-map
```

It has constructs to support data that can be pushed to subscribers and storages (push). Data can be queried from storage and evaluation (remote computation) (pull). Data can also be computed on demand, as shown in Figure 7. It handles dynamic data discovery and adaptive routing with built-in fault-tolerant and load balancing.

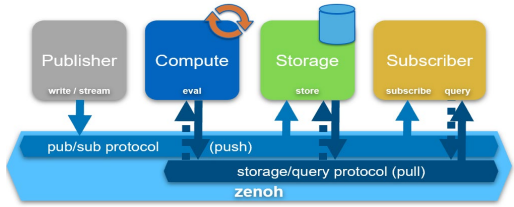

Figure 7 Zenoh Mechanisms

The defined APIs are simple to understand and use. They operate on resources describe as URI, where the resource URI can be bind to either data resources or compute resources. A few simple code snippets shown below (using zenoh syntax):

**Publish :**
```
ws = Zenoh.login().workspace()
ws.put('/demo/hello','Hello world')
ws.put('/city/road/traffic_light', "Red")
```

**Subscribe :**
```
ws = Zenoh.login().workspace()
ws.subscribe('/demo/**', lambda data:
print('received {}'.format(data)))
```

**Query :**
```
ws = Zenoh.login().workspace()
result = ws.get('/demo/hello?(name=World)')
```

*2) The Architecture*

The zenoh data router architecture is shown in Figure 8, it is composed of two components:

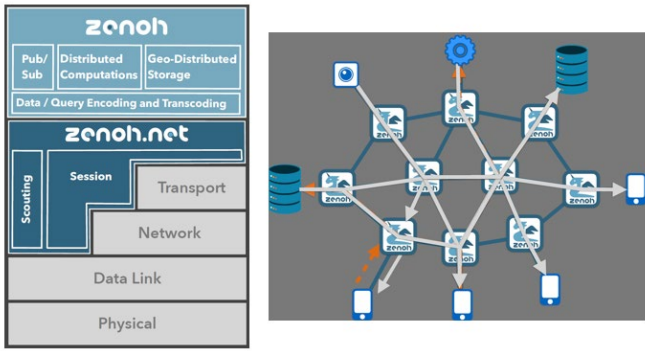

Figure 8 Zenoh Architecture

- Upper layer – zenoh provides high-level APIs for pub/sub and distributed queries, data representation transcoding and implementation of geo-distributed storage and distributed computed values. It defines a series of supported data encoding schemes, such as JSON, Properties, Relational, Raw, etc., along with transcoding. It defines a canonical query syntax based on URIs syntax.

- Lower layer – zenoh.net implements zenoh protocol and data discovery and routing across WAN. The protocol is a wire/power/memory efficient protocol that provides connectivity to extremely constrained targets. It supports push and pull pub/sub along with distributed queries, it supports peer-to-peer and routed communication and it supports ordered reliable data delivery and fragmentation.

*3) DDSF Use Cases*

Efficiently collecting and sharing data among vehicles, edges and cloud enables many new use cases, some examples follow:

- Smart Traffic Light: Assumption that the traffic lights in smart road infrastructure have the built-in data publishing capabilities. When the vehicle enters the area near the traffic light, the vehicle subscribes to the traffic light data such that the vehicle knows the traffic light status, and the ADS algorithm can predict accurately the traffic signal changes and make better decisions.

- Distributed ADS workload: In a recent study [8], it shows some of the AD tasks could be shifted to the external servers, e.g. smart road infrastructure cloud. This so-called edge-cloud computing model for AVs can extend vehicle's computation powers and storage resources. Contrary to conventional wisdom, by moving some heavy workloads to the edge, the AV reduces the execution time and deadline miss ratio despite the added latencies caused by the communications, compared to in-vehicle only computing model.

It is worth noting that traditional pub/sub middleware that are adopted within the vehicle, such as SOME/IP [9] (Scalable service-Oriented Middleware over IP) or DDS [10] (Data Distribution Service), lack remote computation to pre-process raw data, and query capabilities to reduce the data volumes. Another challenge for SOME/IP and DDS is communications across edges and clouds for both north-south and east-west traffics. The distributed data query capability is especially important in vehicle applications since large amount of data are constantly generated by the many vehicle sensors. It's just neither feasible nor efficient for a vehicle to stream all the data out.

### C. DHCF- Distributed Heterogeneous Computing Framework

For autonomous driving there are basically three major application modules: perception, planning and control, for which ML/DL algorithms have been widely used. DL for the perception module it's relatively mature and most of the DL network models are based on convolution neural network. The majority of the perception tasks are done today either in GPU or NPU. However, DL for the planning module it's still an open research area given the complex multi-agent interactive environments. The industry as a whole needs algorithm breakthroughs in behavioral modeling and decision-making.

This paper is not targeted to do the deep dive of AV algorithms, instead we present a dynamic computing framework that can facilitate the AV environment simulation, training and execution. Here are two typical cases to drive the need for DHCF:

1) Deep reinforcement learning has been extremely successful in games, and it's treated as one type of promising algorithms for AV path planning. Different from normal ML/DL framework like TENSORFLOW, PYTORCH etc and task-synchronous parallel computing framework like MAPREDUCE, SPARK etc., it requires training, simulation and serving to be running interactively. Multiple agents are interacting with the environment to generate the observations, then the training subtask takes the accumulated observations to improve the "policy" for some epoch, updating all the agents with the latest policy in an incremental manner.

2) Execution flows in AV to handle heterogeneous computing tasks. For example, the main task running in the CPU would get the camera image from the sensor driver, it then executes one or more perception DNNs tasks in either GPU or NPU in parallel. Those tasks can be dynamic object detection, traffic light detection, road semantic segmentation etc. After getting all the results from the detection tasks, the main task may further do some safety boundary checking and decision. In this case the actor, main task, is running the tasks serially and coordinates all the paralleling detection tasks in GPU or NPU.

DHDF is the framework based on RAY [11] handling such computing scenarios, it is not designed to replace neither the existing ML/DL frameworks nor batch parallel computing framework, instead it works as a scheduler for the heterogeneous computing tasks. For example, TENSORFLOW can use RAY to handle both training and serving in the same context.

### D. *Abstraction for Control, Sensor and V2X Information models*

First, when vehicle E/E architecture moves to be domain or vehicle centralized as shown in figure 1, the vertical 1-1 relationship between controller and actuator will be consolidated into 1 to M, where one logic controller manages multiple actuators or devices, similar to case of the ICT industry where the SDN Controller manages the multi-vendor network devices. Therefore, this is going to drive the needs of open APIs for the controlled actuators and devices.

Second, there are already a large number of sensors installed in the vehicle, and some industry efforts like GENIVI [12] are already working on the signaling format specification. For autonomous driving scenarios there will be multi-vendors environment sensors, and the signaling format specification is critical for downstream data processing.

Third, we expect that 80% vehicles will be equipped with V2X [13]/C-V2X [14] modules in the next 5-10 years. On top of V2X there are many other categories of data and information models that are critical to handle OTA, HD-MAP, ADAS/AV assisted, transportation traffic efficiency and optimization cases (figure 9).

The abstraction layer for control, sensors and V2X are extremely important for data processing, message exchange and control-by-wire for both critical safety and service applications.

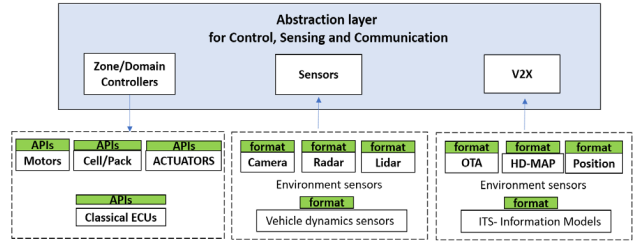

Figure 9- VEC-OF Abstraction Layer in Vehicle

### V. VEC-OF IMPLEMENTATION CONSIDERATION AND CASE WORK FLOWS

As mentioned before VEC-OF leverages some existing open source frameworks as a starting point, then will do the incremental extensions driven through the business cases. Our intention is to lay out one basic software architecture with open interfaces and related data & information models. Adopting open source based implementations we believe this can help accelerate the industry to reach the consensus on the specifications in architecture, interfaces, models and data formats.

In VEC-OF we use KubeEdge as the foundation of OTA, utilize RAY for distributed computing, leverage DDS/Zenoh for distributed data sharing framework, build traffic scenarios and simulations by Scenic [15] and Carla [16]. For details of those existing frameworks and tools please refer to their documentations and implementations directly.

In the appendix section we show four business cases to further explain the capabilities and workflows of VEC-OF.

### VI. CONCLUSION

VEC-OF is designed to address new business requirements in critical areas such as software management, data sourcing mechanism and heterogeneous computing for software defined mobility in-vehicle, across vehicle, edge and cloud, where AUTOSAR might fall short. The transformation journey for the C/C architecture and software platform is very challenging. OEMs and new entrants can take MaaS as an entry point for their digital and software transformation opportunities. As the first industry architecture proposal for the fundamental infrastructure software framework in cooperative connected and autonomous driving domains, VEC-OF helps industries with one open framework that facilitates the collaboration effort in defining architecture, interfaces, models, standards and enables an open software defined ecosystem.

## APPENDIX:

Business cases drive the incrementally development of VEC-OF:

*1) OTA for ECUs and Applications*

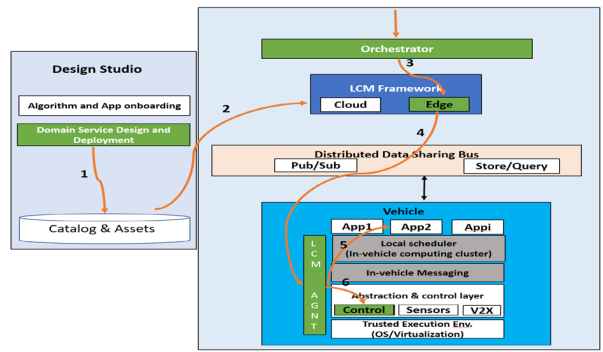

**Message flows:**

1. YAML files are created and stored through service design & deployment portal to the repository along image files

2. The distribution module will deliver the software images to Edge "controller"

3. When the OTA update request comes in, edge controller is invoked to handle the OTA update

4. Edge controller sets up the authentication and encryption channel to the "agent" running in Vehicle side

5. OTA agent handles the update to all the modules running in ECUs and computers. For the containerized apps agent natively handle the software update

6. For ECU update, like the motor controller case, the motor controller will provide the update API, controller handles the update itself

*2) Data sharing for cooperative driving and operation intelligences*

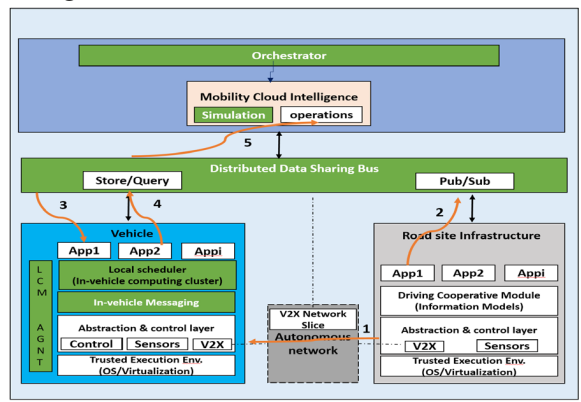

**Message flows:**

1. Road site infrastructure sets up the latency guaranteed service through V2X network service api

2. Infrastructure App processes the road side sensors data and sends the structure semantic sensing data through the distributed data sharing bus

3. Autonomous driving module App1 running in vehicle gets the real-time streaming data from the bus, then does the sensor fusion for next phase path planning

4. Vehicle App2 stores the operational data locally and publishes it to data sharing bus

5. The operation intelligent app "query" the operational data from vehicles offline

*3) Environment and Traffic Simulation for AV*

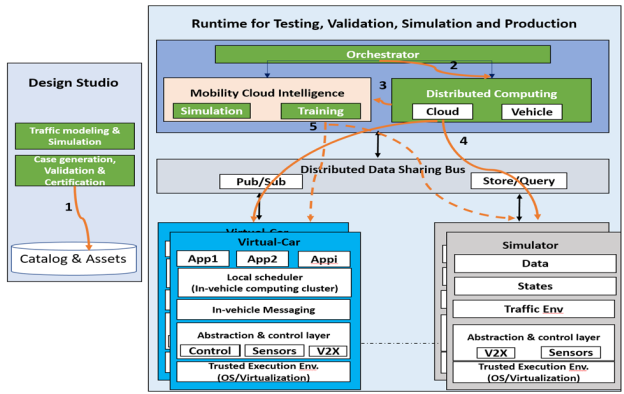

**Message flows:**

1. Traffic environment is programmed in design studio
2. The testing and simulation tasks are processed by the Distributed Computing scheduler
3. Computing scheduler schedule the tasks in the cloud cluster in a heterogeneous computing environment
4. Car simulation and testing environment are created
5. In Training-Severing-Simulation all in one case (RL for AV for example) parallel simulators will be created for high performance and efficient training & validation

*4) CloudRobot for Mobility Devops as a Service*

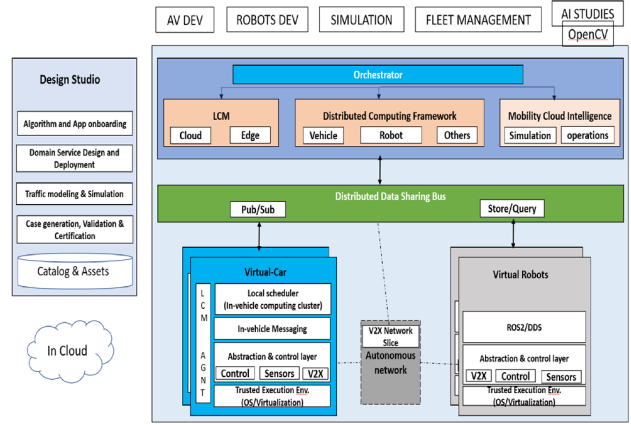

VEC-OF plus others framework like ROS, OpenCV etc. work as development platform to facilitate the industry application developments and researches on AI algorithm